\documentclass[reprint, amsmath, amssymb, groupaddress, aps, prb]{revtex4-1}
\usepackage{graphicx}
\usepackage{amsmath}
\begin{document}
\title{Single-Layer Antiferromagnetic Semiconductor CoS$_2$ with the Pentagonal Structure}
\author{Lei Liu}
\author{Immanuella Kankam}
\author{Houlong L. Zhuang}
\email{zhuanghl@asu.edu}
\affiliation{School for Engineering of Matter Transport and Energy, Arizona State University, Tempe, AZ 85287, USA}
\date{\today}
\begin{abstract}
Structure-property relationships have always been guiding principles in discovering new materials. Here we explore the relationships to discover novel two-dimensional (2D) materials with the goal of identifying 2D magnetic semiconductors for spintronics applications. In particular, we report a density functional theory + $U$ study of single-layer antiferromagnetic (AFM) semiconductor CoS$_2$ with the pentagonal structure forming the so-called Cairo Tessellation. We find that this single-layer magnet exhibits an indirect bandgap of 1.06 eV with light electron and hole effective masses of 0.03 and 0.10 $m_0$, respectively, which may lead to high carrier mobilities. The hybrid density functional theory calculations correct the bandgap to 2.24 eV. We also compute the magnetocrystalline anisotropy energy (MAE), showing that the easy axis of the AFM ordering is out of plane with a sizable MAE of 153 $\mu$eV per Co ion. We further calculate the magnon frequencies at different spin-spiral vectors, based on which we estimate the N$\acute{e}$el temperatures to be 20.4 and 13.3 K using the mean field and random phase approximations, respectively. We then apply biaxial strains to tune the bandgap of single-layer pentagonal CoS$_2$. We find that the energy difference between the ferromagnetic and AFM structures strongly depends on the biaxial strain, but the ground state remains the AFM ordering. Although the low critical temperature prohibits the magnetic applications of single-layer pentagonal CoS$_2$ at room temperature, the excellent electrical properties may find this novel single-layer semiconductor applications in optoelectronic nanodevices.
\end{abstract}
\maketitle
\section{Introduction}
In the book {\it Thank You for Being Late} by the renowned Pulitzer Prize winner Thomas Friedman,\cite{friedman2017thank} Moore's law representing the technology is deemed as one of the three largest forces\textemdash the other two being the market referring to globalization and mother nature alluding to climate change and biodiversity loss\textemdash which are changing our planet at a breathtaking pace. For more than four decades, Moore's law has accurately predicted that the number of transistors in an integrated circuit doubles every two years. But as the size of the circuit keeps shrinking, it is expected that Moore's law will come to an end in the foreseeable future. A road map for addressing these challenges is to make use of magnetic materials at the nanoscale by utilizing spin instead of charge properties of electrons.\cite{schapers2017semiconductor} As such, atomically thin two-dimensional (2D) magnets are promising candidate materials to keep Moore's law alive.  

2D magnets are critical building components for future generations of computers that rely on the usage of spin field-effect transistors. Experimental efforts have been expended to obtain 2D magnets. 2D magnets such as CrI$_3$,\cite{huang2017layer} Cr$_2$Ge$_2$Te$_6$,\cite{gong2017discovery} and Fe$_2$GeTe$_3$\cite{fei2018two} have recently been obtained in experiments. It is expected that such a list of 2D magnets will grow longer. 

Most of the above-mentioned, cutting-edge 2D magnets adopt hexagonal structures. Recently, 2D materials with pentagonal, corrugated structures have attracted intense attention due to a wealth of exotic physical properties and potential applications that are associated with the pentagonal structure. For example, Sn$X_2$ ($X$= S, Se, and Te) consisting entirely of pentagonal rings is quantum spin Hall insulator that produces sizable nontrivial gaps and maintain robust band topology.\cite{ma2016room} 2D pentagonal Sn$X_2$ also present promising applications in low-power-consuming electronic devices. More recently, Akinola {\it et al.} show that 2D pentagonal PdSe$_2$ possess similar indirect and direct bandgaps (1.30 and 1.43 eV, respectively), useful for optoelectronic applications. Moreover, 2D pentagonal PdSe$_2$ exhibits high electron mobility and air stability, making it a candidate for field-effect transistors applications.\cite{oyedele2017pdse2, chow2017high}

Other examples of recently predicted 2D pentagonal materials with the $AB_2$ formula include B$_2$C,\cite{li2015flexible} B$_2$N$_4$,\cite{yagmurcukardes2015pentagonal}  B$_4$N$_2$,\cite{yagmurcukardes2015pentagonal} CN$_2$,\cite{zhang2016beyond} SiC$_2$,\cite{liu2013novel,lopez2015sigma,liu2016disparate} and SiN$_2$.\cite{liu2016disparate} The structure-property relationships are also manifested in these 2D materials with pentagonal, buckling structures. For instance, 2D pentagonal B$_2$C possesses in-plane structural flexibility. It transforms from a buckled structure to a planar structure under biaxial tensile strains, with the bandgaps reduced from 2.28 eV to 0.06 eV, allowing for potential applications in flexible and stretchable electronics.\cite{li2015flexible} In addition to the $AB_2$ formula, 2D pentagonal materials also assume other chemical formulae such as $AB$ and $AB_3$. Transition-metal borides/carbides (TMB/Cs) with the chemical formula of $AB$ have been intensively studied. Shao {\it et al.} predicted pentagonal TMB/Cs (TaB, WB, ZrC, HfC, and TaC) to be thermodynamically stable, among which WB and ZrC show substantial performance for the hydrogen evolution reaction.\cite{shao2018exploring}

\begin{figure}
  \includegraphics[width=8cm]{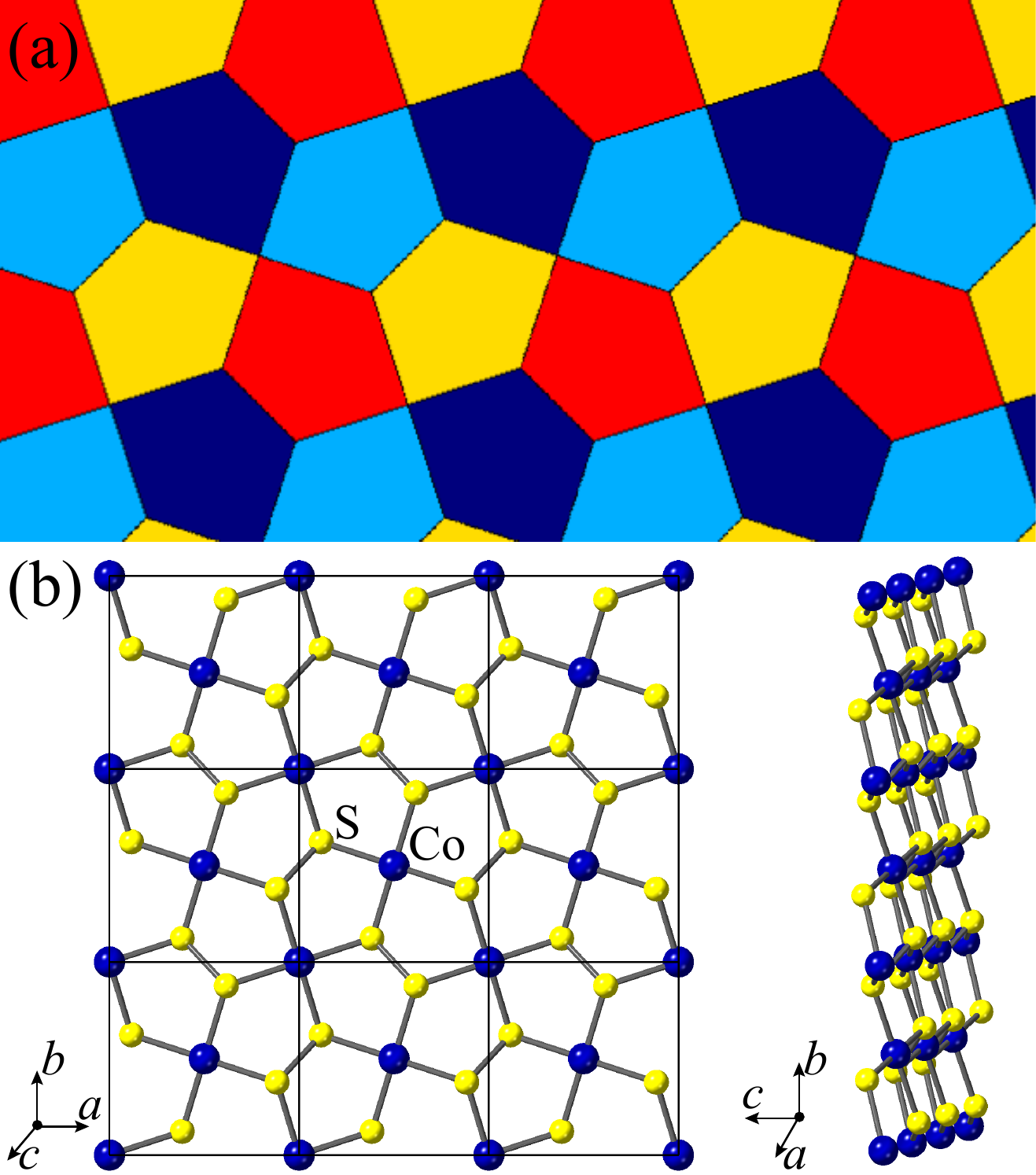}
  \caption{(a) A sketch of the Cairo tessellation formed from type 2 pentagons. (b) Top and side views of a 3 $\times$ 3 supercell of single-layer CoS$_2$ adopting the pentagonal structure.}
  \label{fig:structure}
\end{figure}

Inspired by the geometries of the existing 15 types of convex pentagons that can tessellate a plane without creating a gap or overlap, we recently combined these pentagonal geometries and density functional theory (DFT) calculations to predict novel 2D materials.\cite{liu2018,liu2018encoding} For example, we discovered a hidden pattern of pentagons called the Cairo tessellation in a group of bulk materials with the pyrite structure, having a chemical formula of $XY_2$ and the space group $pa\bar3$. Figure~\ref{fig:structure} illustrates the Cairo tessellation resulted from tessellating type 2 pentagons in a plane. We used single-layer PtP$_2$ as an example and predicted it to exhibit a completely planar pentagonal structure with a direct band gap.\cite{liu2018ptp} In this work, we aim to computationally identify a single-layer pentagonal material with a magnetic ordering suitable for spintronics applications. Because CoS$_2$ has the same crystal structure as PtP$_2$, and a combination of Co with another element could lead to a 2D magnet, single-layer pentagonal CoS$_2$ becomes a natural candidate for this theoretical study. 
\section{Methods}
 We perform the DFT calculations using the Vienna {\it Ab-initio} Simulation Package (VASP, version 5.4.4).\cite{Kresse96p11169} We also use the Perdew-Burke-Ernzerhof (PBE) functional for approximating the exchange-correlation interactions.\cite{Perdew96p3865} The standard PBE version of the potential datasets for Co and S generated based on the projector-augmented wave (PAW) method are used for describing the electron-ion interactions.\cite{Bloechl94p17953,Kresse99p1758} We choose the plane waves with the kinetic cutoff energy below 550~eV to approximate the total electron wave function. Figure 1(b) illustrates a 3~$\times$~3~$\times$~1 supercell of single-layer CoS$_2$ and each unit cell consists of two formula units. We use a $\Gamma$-centered $12~\times~12~\times~1$ $k$-point grid for the integration in the reciprocal space.\cite{PhysRevB.13.5188} We additionally use an effective $U$ parameter $U_\mathrm{eff}$ of 3.32 eV with the Dudarev method\cite{PhysRevB.57.1505} to treat the $d$ orbitals of Co atoms. This $U_\mathrm{eff}$ parameter is taken from Ref.~\onlinecite{PhysRevB.73.195107} and used in the Materials Project for many compounds containing Co.\cite{Jain2013} We create a surface supercell model of single-layer CoS$_2$, which is essentially a single-layer surface slab of the CoS$_2$ (001) surface. The vacuum spacing of the surface slab is set to 18.0~\AA~to avoid image interactions between the monolayers. VASP fully optimizes the in-plane lattice constants along with atomic coordinates in all the three directions until the residual Hellman-Feynman forces are smaller than 0.01 eV/\AA.
\section{Results and Discussion}
\begin{table}[b]
  \caption{Optimized in-plane lattice constants $a$ and $b$ (in \AA), energies with reference to LS-AFM (low spin and antiferromagnetic) state $\Delta E$ ( in meV per formula unit), band gaps $E_\mathrm{g}$ (in eV), and magnetic moments $M$ (in $\mu_\mathrm{B}$ per Co ion)  of single-layer CoS$_2$ with different spin states. HS: high spin;  IS: intermediate spin;  LS: low spin; AFM: antiferromagnetic; FM: ferromagnetic; NM: non-magnetic. All these results are obtained from the PBE + $U$ ($U_\mathrm{eff}$ = 3.32 eV) calculations.}
  \begin{ruledtabular}
    \begin{center}
      \begin{tabular}{cccccc}
            Spin state & $\Delta E$ &$a$  &$b$ & $m$ & $E_\mathrm{g}$  \\
            \hline
HS-AFM&101.17&5.78&5.78&0.00&1.41\\
HS-FM&252.91&5.75&5.76&2.34&Metal\\
IS-AFM&0.00&5.34&5.43&0.00&1.06\\
IS-FM&11.99&5.33&5.43&1.00&1.97\\
LS-AFM&0.00&5.34&5.43&0.00&1.06\\
LS-FM&11.99&5.33&5.43&1.00&1.97\\
NM&663.45&5.40&5.40&0.00&Metal\\
      \end{tabular}
    \end{center}
  \end{ruledtabular}
  \label{summary}
\end{table}
We first determine the ground-state magnetic ordering of single-layer pentagonal CoS$_2$. Because Co ions in this single-layer material adopt the $d^5$ configuration in a nearly square planar crystal field, we consider three possible spin states: high-spin (HS), intermediate-spin (IS), and low-spin (LS). The three spin states are distinguished by setting the initial magnetic moments to 5.0, 3.0, and 1.0 $\mu_B$ ($\mu_B$: Bohr magneton), respectively, in the VASP calculations. For each spin state, we consider two possible magnetic orderings: antiferromagnetic (AFM) and ferromagnetic (FM).  For comparison, we also perform a non-magnetic (NM) calculation, where the magnetic moments of Co ions are not taken into account. Table~\ref{summary} compares the optimized total energies, the in-plane lattice constants, and the total magnetic moments of the six spin states and of the NM state. We observe that the LS AFM state is the ground state. The optimized in-plane geometry is nearly a square with the lattice constant $b$ (5.43~\AA) slightly larger than $a$ (5.34~\AA). The net magnetization is zero and the two Co ions in the unit cell possess anti-parallel magnetic moments with the same magnitude of about 1.0 $\mu_B$. Unlike single-layer PtP$_2$ with a planar Cairo tessellation,\cite{liu2018} single-layer pentagonal CoS$_2$ with the AFM ordering displays a buckled pentagonal structure as shown in Fig.~\ref{fig:structure}(b). Each Co ion is surrounded by four S ions, and the Co-S bond length is 2.20 or 2.22~\AA. The sublayer of Co ions is sandwiched between the top and the bottom sublayers of S ions; the interlayer distance of the Co and S sublayers is 0.54~\AA. The energy of the FM state of the LS configuration is higher than the ground state by around 12.0 meV per formula unit. We also find that the IS AFM and FM states are relaxed to the LS AFM and FM states, respectively. By contrast, the HS AFM and FM states are optimized into the nearly IS state with the magnetic moment of 2.34 $\mu_B$. The NM state exhibits the highest energy, showing the importance of considering spin polarization in the calculations. 

\begin{figure}
  \includegraphics[width=8cm]{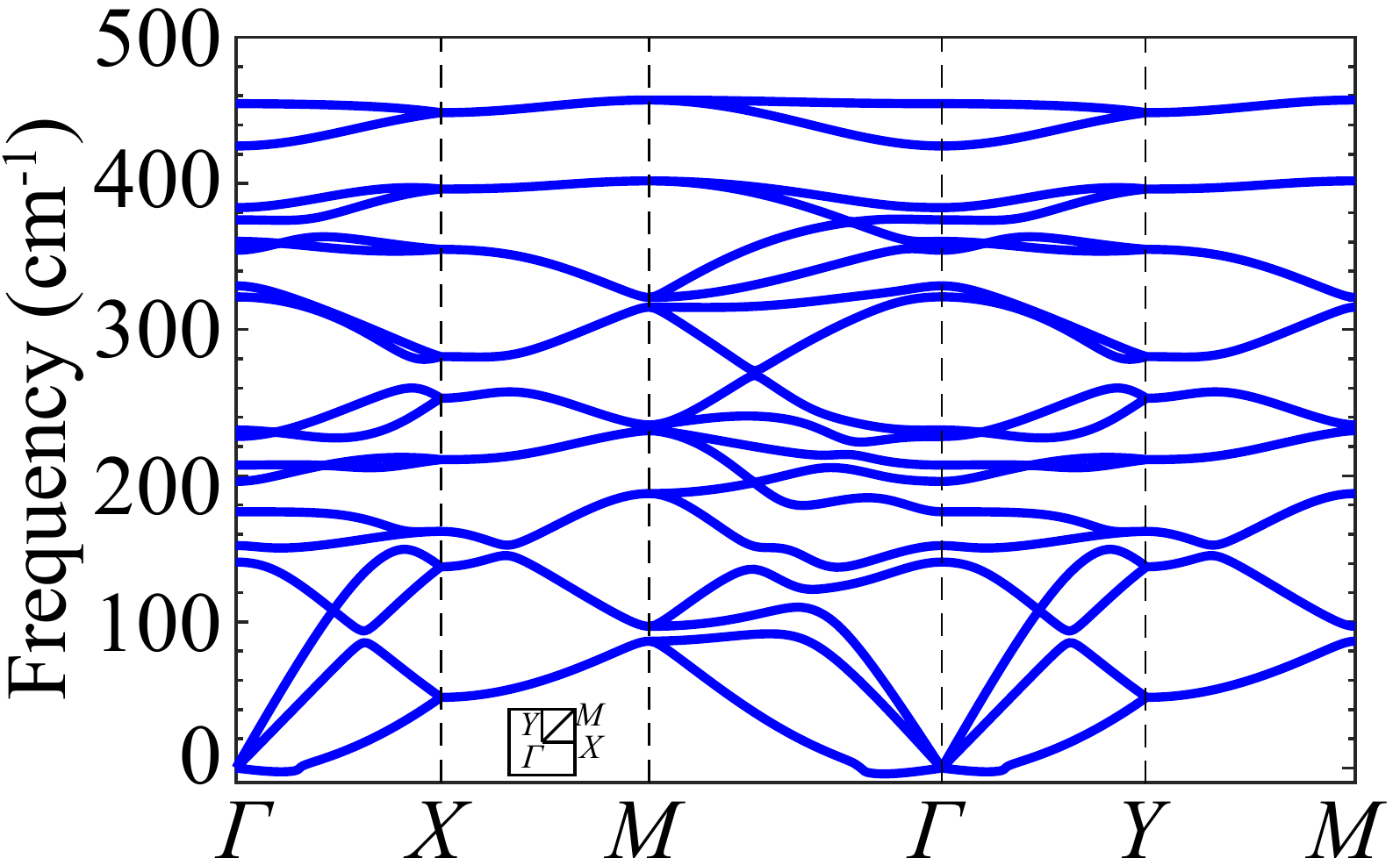}
  \caption{Predicted phonon spectrum of single-layer pentagonal CoS$_2$. The inset rectangle illustrates the first Brillouin Zone with the high-symmetry phonon $q$ points denoted.}
  \label{fig:phonon}
  \end{figure}

To confirm the dynamic stability of single-layer pentagonal CoS$_2$ structure with the AFM ordering, we calculate its phonon spectrum with the force information obtained from VASP calculations for 3~$\times$~3~$\times$~1 supercells followed by post-processing calculations via Phonopy.\cite{phonopy} Figure~\ref{fig:phonon} shows the calculated phonon spectrum. Although there are some negligibly small imaginary frequencies near the $\Gamma$ point due to the translational invariance ($i.e.$, the acoustic sum rule), the absence of imaginary modes in the other $q$ points throughout the Brillouin zone shows the dynamic stability of single-layer pentagonal CoS$_2$. 

\begin{figure}
  \includegraphics[width=8cm]{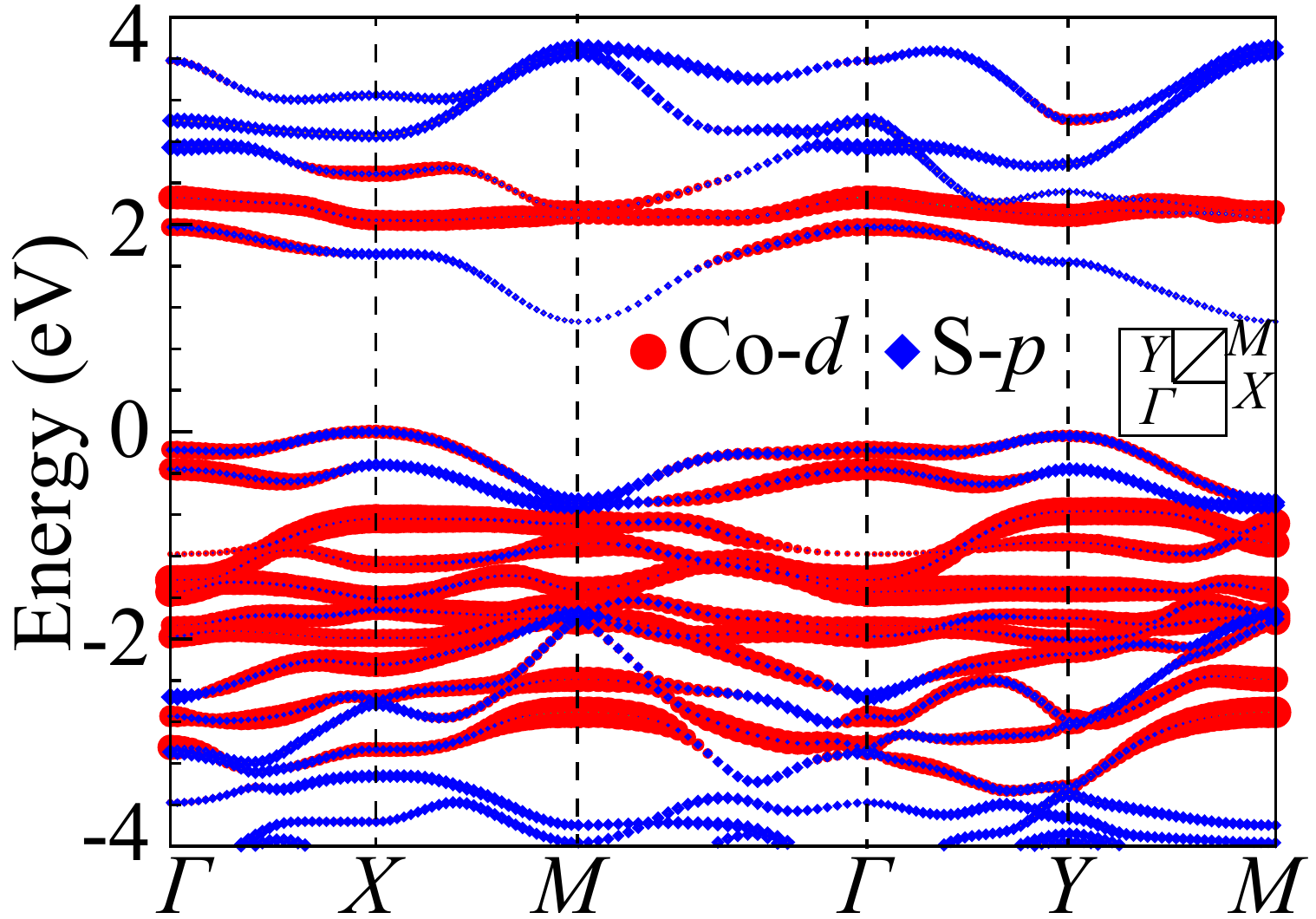}
  \caption{Orbital-resolved band structure of single-layer pentagonal CoS$_2$ calculated with the PBE + $U$ ($U_\mathrm{eff}$ = 3.32 eV) method. The valance band maximum is set to zero.}
  \label{fig:band}
\end{figure}
The pentagonal structure of single-layer CoS$_2$ leads to the orbital-resolved band structure illustrated in Fig.~\ref{fig:band}. As can be seen, single-layer pentagonal CoS$_2$ is a semiconductor with an indirect bandgap of 1.06 eV\textemdash we also compute the bandgaps of the other spin configurations and the band gaps are listed in Table~\ref{summary}. The orbitals at the valence band maximum (VBM) are from the contributions of mixed $d$ orbitals of Co atoms and $p$ orbitals of S atoms. The conduction band minimum (CBM) is located at the $X$ point. The top valence band near the $Y$ point has almost the same energy as the CBM, but slightly smaller. The band dispersions near the CBM and VBM both show parabolic relationships, indicating that the carriers resulted from doping exhibit a two-dimensional electron/hole gas behavior and that the electron/hole effective masses are nearly isotropic. We calculate the electron effective mass $m_e^*$ at the CBM as 0.03 $m_0$ ($m_0$: the electron rest mass) and the hole effective mass $m_h^*$ at the VBM as 0.10 $m_0$. These masses are much lighter and also less anisotropic than the masses in other popular single-layer semiconductors such as molybdenum disulphide ($m_e^*$ = 0.34 $m_0$; $m_h^*$ = 0.46 $m_0$)\cite{cheiwchanchamnangij2012quasiparticle} and black phosphorene ($m_e^* \approx m_h^* \approx 0.30~m_0$).\cite{liu2014phosphorene} Such small effective masses may lead to high carrier mobilities.
  \begin{figure}
  \includegraphics[width=8cm]{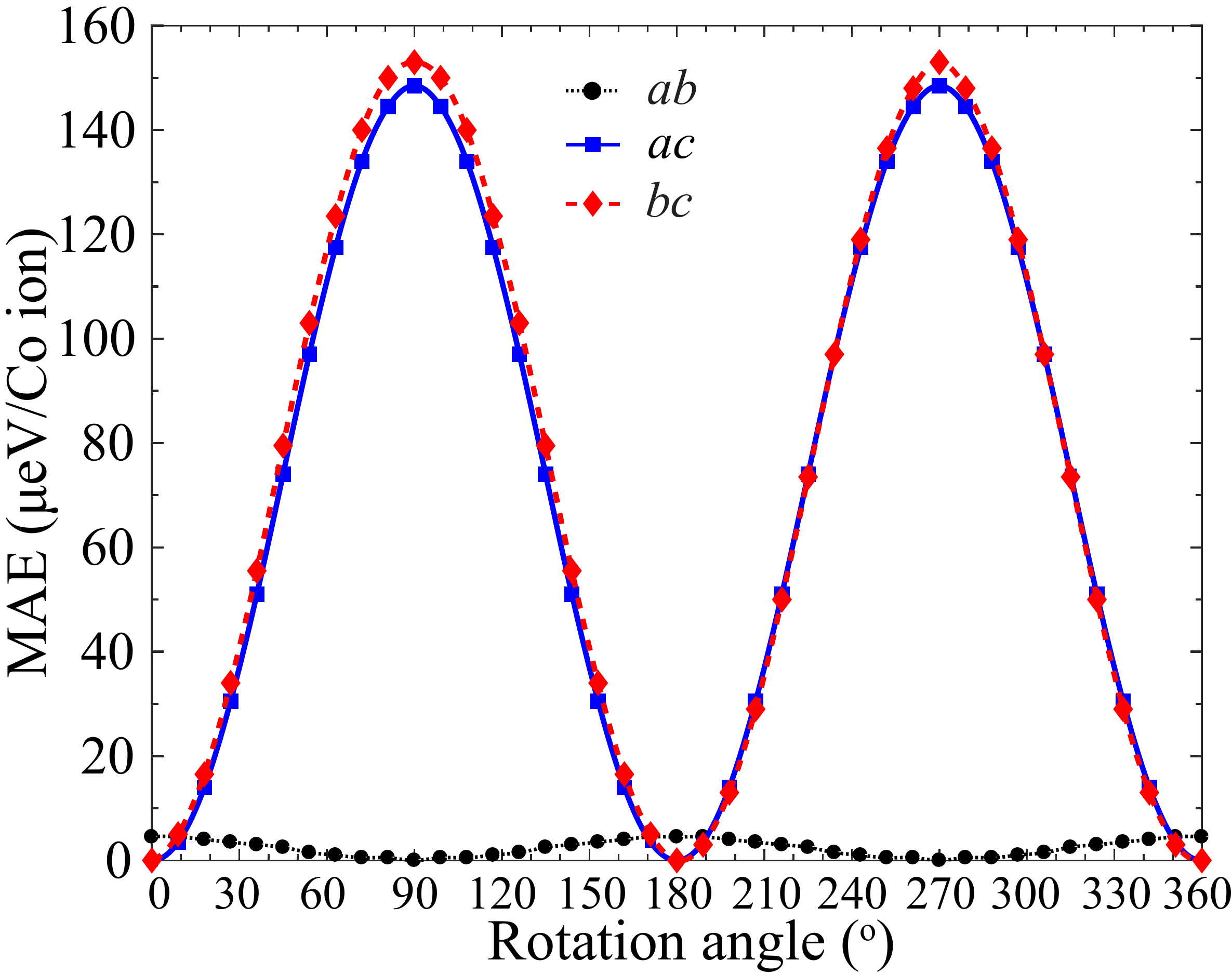}
  \caption{Magnetocrystalline anisotropy energy (MAE) of single-layer pentagonal CoS$_2$ with the spins of the two Co ions antiparallelly oriented in different directions in the $ab$, $bc$, and $ac$ planes.}
  \label{fig:mae}
  \end{figure}
We also use the HSE06 hybrid density functional\cite{heyd2003hybrid} to correct the possibly underestimated bandgap of single-layer pentagonal CoS$_2$ with the PBE + $U$ ($U_\mathrm{eff}$ = 3.32 eV) method. Indeed, the calculated HSE06 bandgap (2.24 eV) is nearly twice as wide as the PBE + $U$ bandgap (1.06 eV). The HSE06 functional also confirms the ground state is the AFM ordering with the energy lower than that of the FM ordering by 8.45 meV per formula unit.   
  
The occurrence of the AFM ordering in single-layer CoS$_2$ follows the Goodenough-Kanamori rules,\cite{PhysRev.100.564,KANAMORI195987} which state that the type of magnetic ordering due to the superexchange interactions between two metal ions bridged by a nonmetal ion depends on the metal-nonmetal-metal bond angle. If this angle is near 90$^\circ$, the resulting magnetic ordering is FM; if the angle is greater than 90$^\circ$ the corresponding ordering is AFM. The calculated Co-S-Co angle is 118.5$^\circ$. We therefore suggest the mechanism for the AFM ordering is due to superexchange interactions, where the Co-Co ions are far away from each other, and the interactions between these two ions must be bridged by the non-metallic S ions. An intuitive physical picture\cite{mabbs2008magnetism}  of this superexchange interactions is that one electron in one of the $d$ orbitals ($e.g.$, spin-up $d_{z^2}$) of a Co ion overlaps with one of the $p$ orbitals ($e.g.$, spin-down $p_z$) of the bridging S ion to form a $\sigma$ bond. This leads to a remaining spin-up electron in the $p_z$ orbital overlapping with the spin-down $d_{z^2}$ orbital in another Co ion. The net effect of these superexchange interactions is the AFM ordering.

To identify the easy axis of single-layer pentagonal CoS$_2$, we compute the magnetocrystalline anisotropy energy (MAE) with the torque method as implemented in VASP.\cite{PhysRevB.54.61} Figure~\ref{fig:mae} displays the variations of the MAEs with the spin orientations in the $ab$, $bc$, and $ac$ planes. In the ac plane, the MAE is almost negligible while the MAEs in the other two planes are similar and much stronger. The zero rotation angle in Fig.~\ref{fig:mae} refers to the spin configuration where the spin axes of the two Co ions are parallel and anti-parallel to the $c$ axis, respectively. The positive MAEs at the other rotation angles therefore show that the easy axis for the AFM ordering of single-layer pentagonal CoS$_2$ is the $c$ axis. The highest MAE (153 $\mu$eV/Co ion) occurs when the spin axes are along the $b$ direction.

\begin{figure}
  \includegraphics[width=8cm]{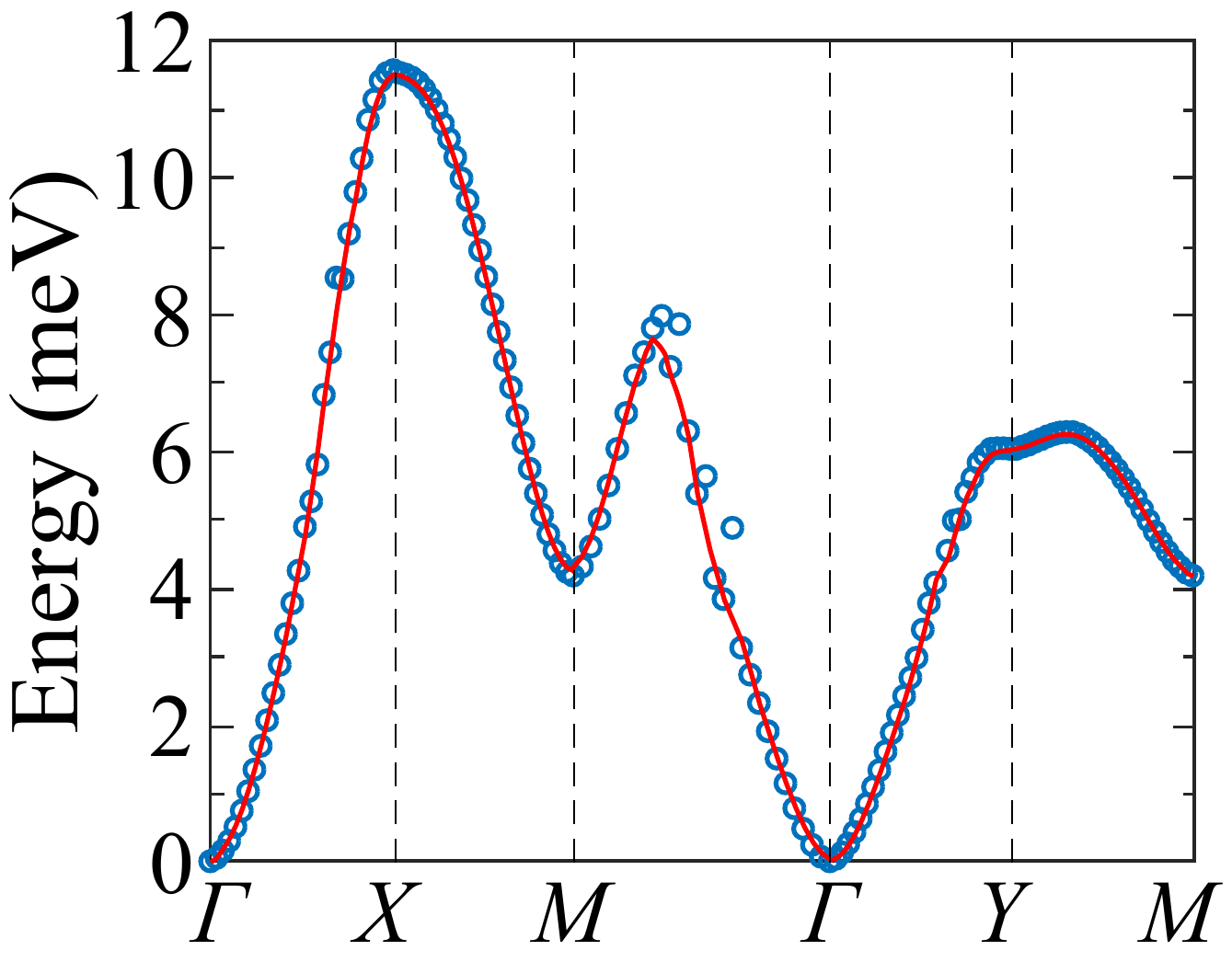}
  \caption{Predicted magnon spectrum of single-layer pentagonal CoS$_2$ using the PBE + $U$ ($U_\mathrm{eff}$ = 3.32 eV) method. The open circles represent our calculated data; the solid red line is used only to aid the view of the magnon dispersions.}
  \label{fig:magnon}
  \end{figure}

We also calculate the magnon spectrum of single-layer pentagonal CoS$_2$ using the frozen magnon method.\cite{PhysRevB.58.293} We set a small polar angle 3.0$^\circ$ as suggested in Ref.~\onlinecite{PhysRevB.84.174425}. Figure~\ref{fig:magnon} shows a magnon spectrum along the high-symmetry $q$ point path. We also use a 5 $\times$ 5 $\times$ 1 $q$-point grid of spin-spiral vectors in the first Brillouin Zone (BZ) to compute the magnon energies at these $q$ vectors. We then calculate the N$\acute{e}$el temperatures via the mean field and random phase approximations ($T_\mathrm{N}^\mathrm{MFA}$ and $T_\mathrm{N}^\mathrm{RPA}$) written as\cite{PhysRevB.84.174425}
\begin{equation}
k_\mathrm{B}T_\mathrm{N}^\mathrm{MFA} = \frac{M}{3}{\bigg[\frac{1}{N}\sum_{\bf q = 0}^{\mathrm{BZ}}{\omega_\mathrm{\bf q}}\bigg]},
\label{eq8}
\end{equation}
and
\begin{equation}
k_\mathrm{B}T_\mathrm{N}^\mathrm{RPA} = \frac{M(N-1)}{3}{\bigg[\sum_{\bf q\neq 0}^{\mathrm{BZ}}\frac{1}{\omega_\mathrm{\bf q}}\bigg]}^{-1},
\label{eq9}
\end{equation}
where $k_\mathrm{B}$ is the Boltzmann constant, and $N$ is the total number of {\bf q} vectors, {\it i.e.,} $N$ = 25. The calculated $k_\mathrm{B}T_\mathrm{N}^\mathrm{MFA}$ and $k_\mathrm{B}T_\mathrm{N}^\mathrm{RPA}$ are 20.4 and 13.3 K, respectively, excluding the practical magnetic applications of single-layer pentagonal CoS$_2$ at ambient conditions. This issue of low critical temperature  seems to be a common challenge faced by 2D magnets ({\it e.g.,} the Curie temperature of CrI$_3$ is merely around 45 K),\cite{huang2017layer} requiring joint experimental and theoretical efforts in the future work.

The low, predicted critical temperatures of single-layer pentagonal CoS$_2$ are due to the weak superexchange interactions between Co ions. To estimate the strength of these superexchange interactions, we use the energy difference approach to derive the exchange integral $J_1$ between the nearest-neighboring Co ions in a unit cell. By mapping the magnetic interactions between Co ions onto the Heisenberg Hamiltonian,\cite{fischer2009exchange} we calculate $J_1$ as 
\begin{equation}
J_1 = \frac{E_\textrm{FM}-E_\textrm{AFM}}{8M^2},
\label{eq7}
\end{equation}
where $M$ denotes the dimensionless magnitude ($M$ = 1.0) of the magnetic moment of a Co ion. $E_\textrm{FM}$ and $E_\textrm{AFM}$ are the energies of CoS$_2$ unit cell with FM and AFM configurations, respectively. The calculated $J_1$ from Eq.~\ref{eq7} and using the energy difference shown in Table~\ref{summary} is 3.01 meV, which is much smaller than that of other predicted 2D magnets such as Co$_2$S$_2$ with the $J_1$ of 58.7 meV.\cite{zhang2017two}

\begin{figure}
 \includegraphics[width=8cm]{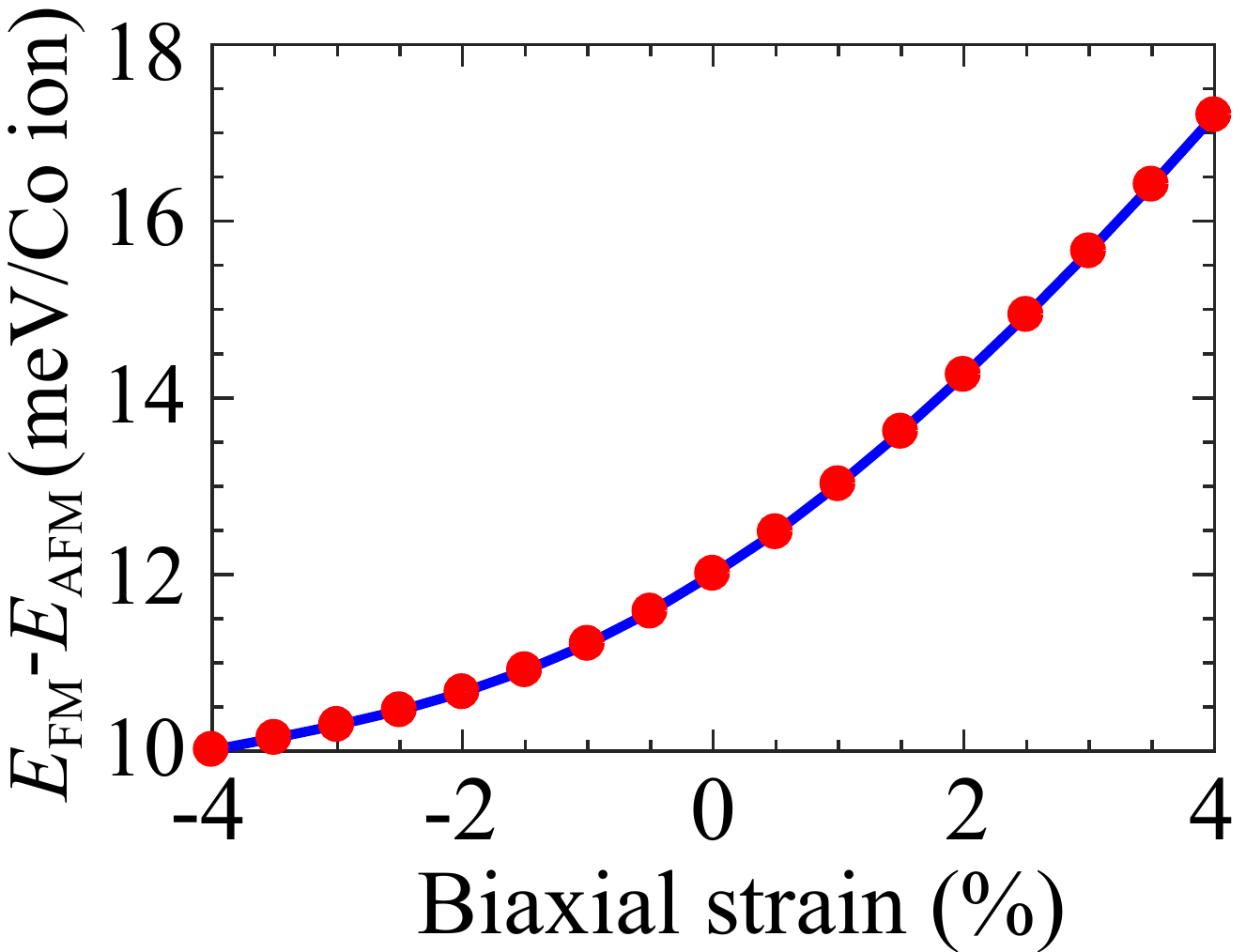}
 \caption{Energy differences between single-layer pentagonal CoS$_2$ with the ferromagnetic and antiferromagnetic orderings under biaxial strains ranging from -4\% to 4\%.}
 \label{fig:deltaE}
 \end{figure}
  
Strain engineering has been widely used to tune the structural and electrical properties of single-layer materials, offering an important degree of flexibility.\cite{wei2011strain,conley2013bandgap, fei2014strain} We apply in-plane biaxial strains ($\epsilon_{aa}$ = $\epsilon_{bb}$) to single-layer pentagonal CoS$_2$. We first examine whether the strains induce a transition in the AFM ordering to the FM ordering. Figure~\ref{fig:deltaE} shows the energy difference of single-layer pentagonal CoS$_2$ with the FM and AFM orderings under the biaxial stains ranging from -4\% to 4\% at an incremental step of 0.5\%. We observe that the energy difference remains positive in the range of applied biaxial strains, showing that the AFM ground state is unaffected by the strains. Furthermore, the energy difference increases with the increasing tensile strains and decreases with the increasing compressive strains. According to this trend, one probably needs to apply an extremely large compressive strain to turn the AFM to the FM ordering.
  \begin{figure}
  \includegraphics[width=8cm]{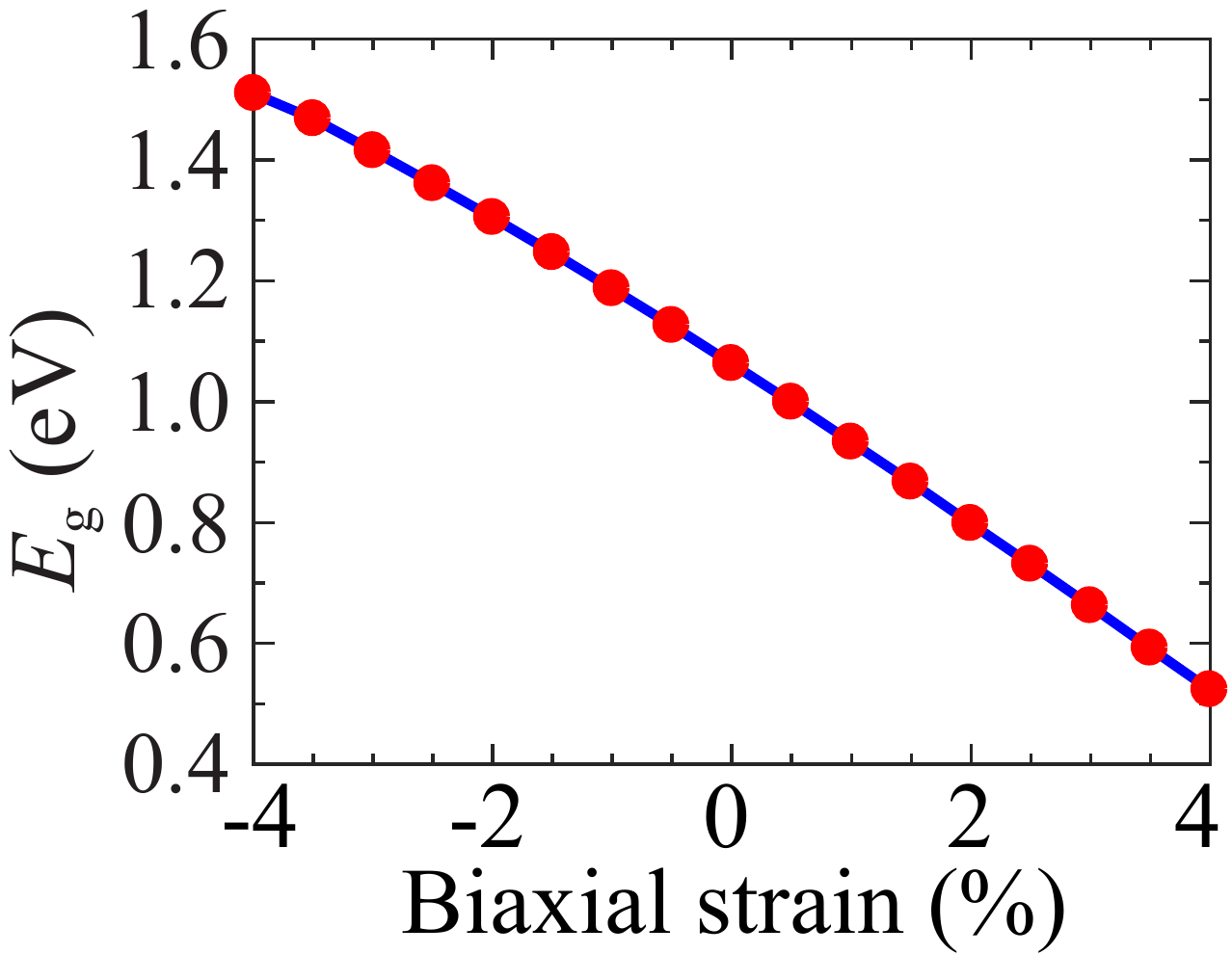}
  \caption{Bandgaps of single-layer pentagonal CoS$_2$ under biaxial strains ranging from -4\% to 4\%.}
  \label{fig:bandgap}
  \end{figure}
\begin{figure}
\includegraphics[width=8cm]{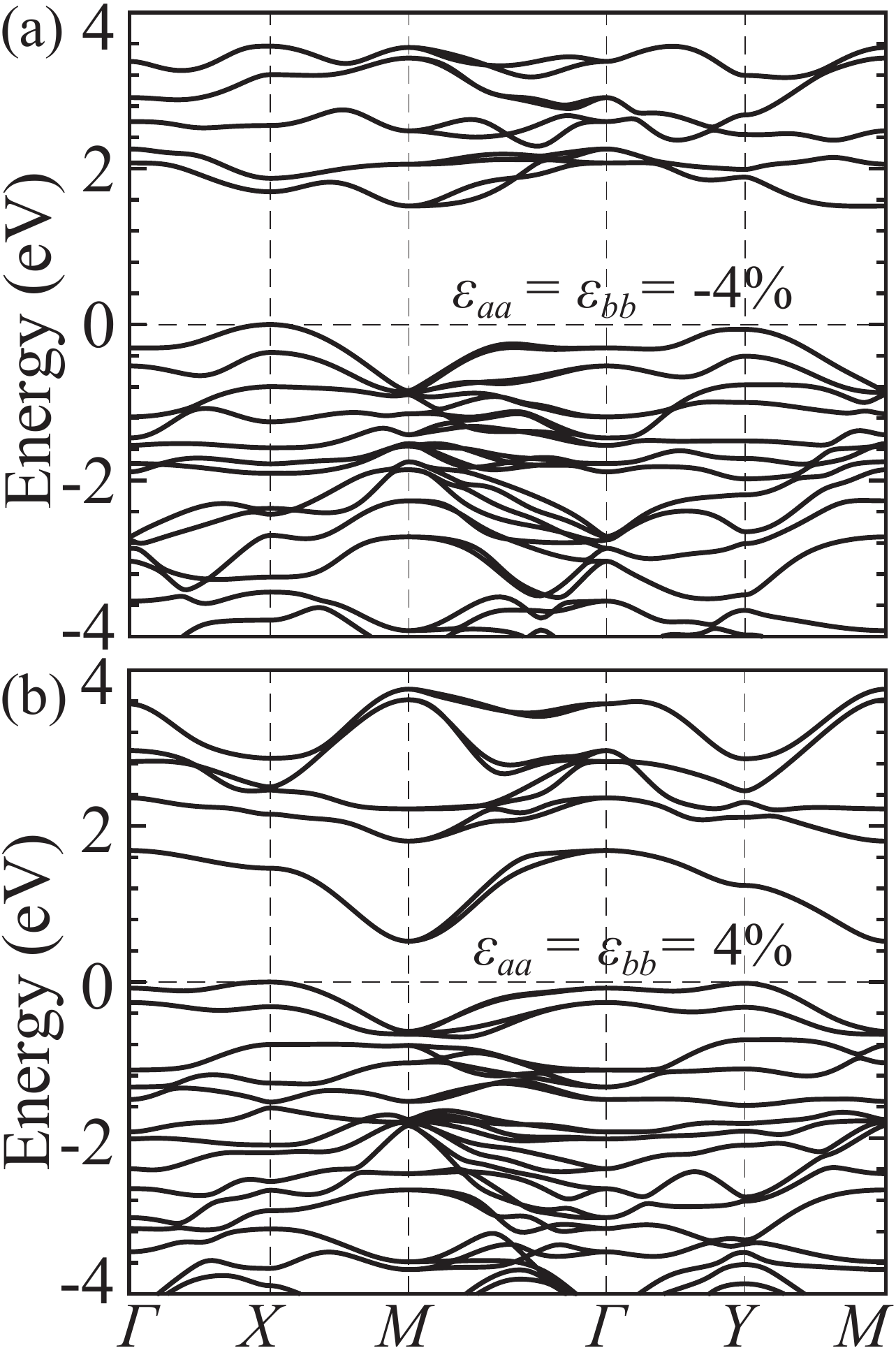}
\caption{Band structures of single-layer pentagonal CoS$_2$ under biaxial strains of (a) -4\% and (b) 4\% calculated with the PBE + $U$ ($U_\mathrm{eff}$ = 3.32 eV) method. The valance band maxima are set to zero.}
\label{fig:bandstructure_strain}
\end{figure}

The bandgap size of single-layer pentagonal CoS$_2$ is significantly affected by the biaxial strains. Figure~\ref{fig:bandgap} shows that the bandgaps decrease almost linearly from 1.51 eV at the maximum compressive strain ($\epsilon_{aa}$ = $\epsilon_{bb}$= -4\%) to 0.52 eV at the maximum tensile strain ($\epsilon_{aa}$ = $\epsilon_{bb}$= 4\%), displaying the tunability of -0.12 eV per 1\% of biaxial strain.  The trend of the bandgaps with the biaxial strains\textemdash Namely, the bandgap decreases with increasing bond lengths\textemdash has been typically observed in single-layer ionic materials such as BN.\cite{qi2012strain,zhuang2012electronic}  The bandgap type of single-layer pentagonal CoS$_2$ under any strain remains indirect , as can be seen from the band structures displayed in Fig.\ref{fig:bandstructure_strain} for the biaxial strains of -4\% and 4\%.

\section{Conclusions}
In summary, we have predicted a new single-layer AFM semiconductor CoS$_2$ via surveying structure-properties relationships for 2D materials. The predicted single-layer material exhibits a pentagonal structure forming the Cairo Tessellation. Our DFT+$U$ calculations also show that single-layer pentagonal CoS$_2$ is an indirect bandgap semiconductor with the bandgap of 1.06 eV. The more accurate hybrid density functional corrects this bandgap to 2.24 eV, within the visible light range, indicating potential energy-related applications of this novel semiconductor. We also found that single-layer pentagonal CoS$_2$ exhibits a sizable MAE with the easy axis along the $c$-axis. We further applied the frozen magnon method to compute magnon frequencies and the N$\acute{e}$el temperatures using the MFA and RPA. The resulting N$\acute{e}$el temperatures are  much smaller than room temperature, make the material not suitable for practical applications. But the remarkable electrical properties of single-layer pentagonal CoS$_2$ such as the bandgap tunable strains and small effective electron/hole masses assures it as a candidate 2D material for optoelectronic nanodevices. 
\begin{acknowledgments}
We thank the Master's Opportunity in Research Engineering (MORE) program and the start-up funds from Arizona State University. This research used computational resources of the Texas Advanced Computing Center under Contracts No.TG-DMR170070. 
\end{acknowledgments}
\bibliography{references}
\end{document}